# Unexpected Hydrophobicity on Self-Assembled Monolayers Terminated with Two Hydrophilic Hydroxyl Groups

Dangxin Mao[1], Xian Wang[1], Yuanyan Wu[1], Zonglin Gu[1], Chunlei Wang[3,4,*] and Yusong Tu[1,2,*]

[1]*College of Physics Science and Technology, Yangzhou University, Jiangsu 225009, China*

[2]*Key Laboratory of Polar Materials and Devices Ministry of Education School of Physics and Electrical Science, East China Normal University, Shanghai 200241, China*

[3]*Division of Interfacial Water and Key Laboratory of Interfacial Physics and Technology, Shanghai Institute of Applied Physics, Chinese Academy of Sciences, Shanghai 201800, China*

[4]*Shanghai Advanced Research Institute, Chinese Academy of Sciences, Shanghai 201210, China*

[*]Email: ystu@yzu.edu.cn; ystu@clpm.ecnu.edu.cn; wangchunlei@zjlab.org.cn

**Current major approaches to access surface hydrophobicity include directly introducing hydrophobic nonpolar groups/molecules into surface or elaborately fabricating surface roughness. Here, for the first time, molecular dynamics simulations show an unexpected hydrophobicity with a contact angle of 82 º on a flexible self-assembled monolayer terminated only with two hydrophilic OH groups ($(OH)_2$-SAM). This hydrophobicity is attributed to the formation of a hexagonal-ice-like H-bonding structure in the OH matrix of $(OH)_2$-SAM, which sharply reduces the hydrogen bonds between surface and water molecules above. The unique simple interface presented here offers a significant molecular-level platform for examining the bio-interfacial interactions ranging from biomolecules binding to cell adhesion.**

Interfacial wettability is of fundamental importance, and various surface-modified functional groups and their surface structures, e.g., self-assembled monolayers (SAM), have been extensively investigated and designed to regulate surface wettability for various given applications [1,2]. Generally, polar functional groups or molecules (e.g., carboxyl (COOH) and hydroxyl (OH)) hold high affinity with water through polar interactions (e.g., hydrogen bonding (H-bond)) and thus are generally thought to assemble into a hydrophilic surface, and vice versa [3-7]. In the past decades of research, there have been two major approaches to access the hydrophobicity on hydrophilic surfaces, including directly introducing nonpolar groups/molecules to form heterogeneously mixed surfaces [8-12] and elaborately fabricating specific surface roughness with micro-/nano- structures based on capillary effect (Wenzel and Cassie–Baxter theories) [13-15]. Whether mixing hydrophobic groups or fabricating surface roughness, the approaches to access surface hydrophobicity are still under the framework of directly synthesizing hydrophobic characters onto surfaces. However, up to now, there is very few reports on how to access hydrophobicity on surfaces constituted by only utilizing hydrophilic groups/molecules. Our previous study has showed the hydrophobicity enhancement with a contact angle of ~34° on the self-assembled monolayers (SAM) terminated only with hydrophilic carboxyl groups (COOH-SAMs) [16].

In this study, we show an unexpected hydrophobicity with a water contact angle of 82° on a flexible SAM terminated only with two hydrophilic OH groups ((OH)$_2$-SAM) at room temperature. This hydrophobicity is attributed to the formation of a hexagonal-ice-like H-bonding structure in the OH matrix that presents not only the oxygen-coordinate assignment consistent with the structure along the basal face of ice I$_h$ but also no dangling H atom of OH groups out of the surface, which sharply reduces the number of H-bonds between the surface and water molecules above. Moreover, within loosely packing densities of alkyl chains, as compensation for the loosened H-bonding structure, water molecules are found to be embedded into the OH matrix to form the embedded-water-OH composite structure on (OH)$_2$-SAM, and the maintained H-bonding network in the composite structure still results in hydrophobicity enhancement compared with the super hydrophilicity on (OH)$_2$-SAM with dense packing densities.

The (OH)$_2$-SAMs were constructed with five-carbon long alkyl chains, which have grafted to one end with two OH groups exposed to water (Fig. 1(a)). The other end is attached to a layer of model atoms restrained by harmonic springs at locations consistent with those on a {111} facet of a face-centered cubic (FCC) lattice. We varied the packing density ($\Sigma$) from 2.0 nm$^{-2}$ to 6.5 nm$^{-2}$. Molecular dynamics (MD) simulations were carried out to analyze contact angles of water droplets on (OH)$_2$-SAMs. This method to calculate the contact angles follows our previous work (also see PS3 in Supporting Information) [16,17]. Initially, the (OH)$_2$-SAM surface along x-y plane was covered completely by a thin slab of water with a thickness of ~0.3 nm. The simulation time was at least 100 ns to ensure the systems to reach thermodynamic equilibrium and the data of the last 20 ns were collected for analyses. The periodic boundary conditions were applied in all directions for all calculations. MD simulations were performed using a time step of 1.0 fs with GROMACS 5.0.6 [18] in NVT ensemble with a velocity-rescaling thermostat at a temperature of 300 K. The OPLS-AA force field [19] was used for (OH)$_2$-SAMs. The SPC/E model [20] was used for the water molecules. The Particle-Mesh Ewald (PME) [21] method was adopted for the long-range electrostatic interactions with a cutoff of 1.2 nm, whereas a 1.2 nm cutoff was applied to the van der Waals interactions.

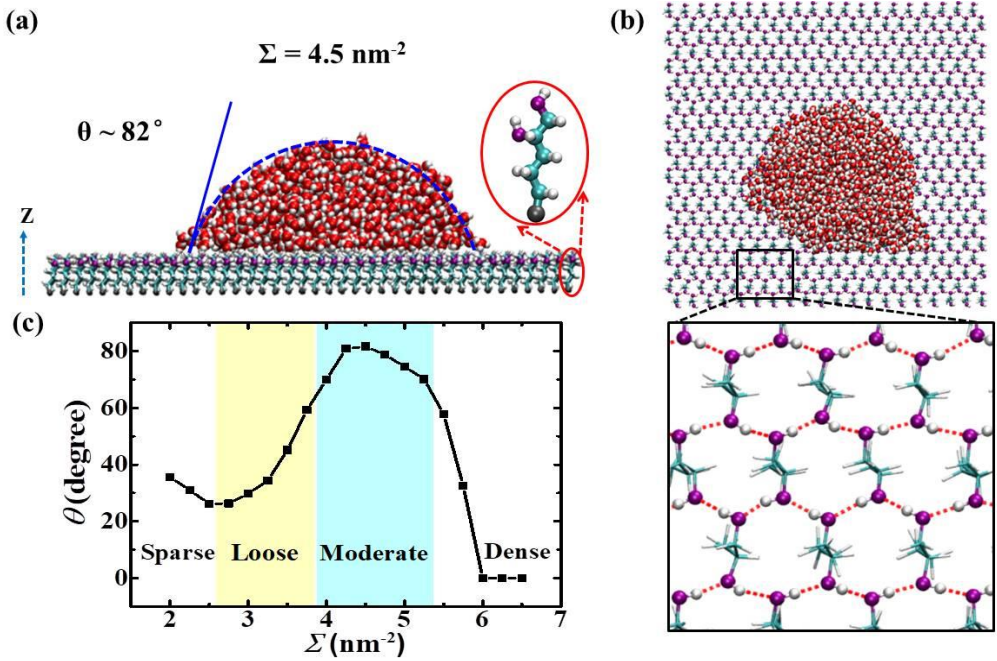

**FIG 1.** (a) The alkyl chain terminated with two hydroxyl (OH) groups and a side view snapshot of water droplet on (OH)$_2$-SAM with a packing density of $\Sigma = 4.5$ nm$^{-2}$ (Model atoms, gray; hydroxyl groups, purple and white; water, red and white; alkyl chains, cyan and white). The water droplet displays the contact angle of 82°. (b) Top view snapshot (top) of the subfigure (bottom) with an enlarged region representing the hexagonal-ice-like H-bond structure on (OH)$_2$-SAM. The H-bonds are shown as red dashed lines. (c) Contact angle $\theta$ of water droplets on (OH)$_2$-SAMs as a function of $\Sigma$. The loose and moderate $\Sigma$ ranges are marked in yellow and blue, respectively.

Unexpectedly, surface hydrophobicity was observed on the (OH)$_2$-SAM that was conventionally regarded as hydrophilic. Figure 1(a) presents the typical snapshot of water droplet on (OH)$_2$-SAM with the packing density $\Sigma = 4.5$ nm$^{-2}$, and the contact angle of the water droplet is up to 82°, representing the hydrophobicity of (OH)$_2$-SAM [1,22,23]. Further, Fig. 1(b) displays an integrated H-bonding structure in the OH matrix of (OH)$_2$-SAM. We observed that, the first (second) OH group of each alkyl chain not only offers its terminal H atom acting as a donor and but also accepts a H atom (reflecting an acceptor) separately H-bonding with the second (first) OH groups of its two neighboring alkyl chains. Such oxygen-coordinate arrangement of the OH groups represents a hexagonal-ice-like structure consistent with the structure along the basal face of ice I$_h$ [24]. Also, it is worth noting that no dangling hydrogen atom appears out of the surface, since the OH groups are analogous to half a water molecule and covalently bonded to alkyl chains. For the (OH)$_2$-SAM of $\Sigma = 4.5$ nm$^{-2}$, the average number of H-bonds in the OH matrix is ~8.8 nm$^{-2}$ (the saturation value is ~9.0 nm$^{-2}$ at $\Sigma = 4.5$ nm$^{-2}$). Within a wide range of $\Sigma$ (3.9 nm$^{-2}$ – 5.4 nm$^{-2}$, marked as *moderate range*), the average H-bond numbers are still larger than ~6.3 nm$^{-2}$ (see Fig. 3), suggesting that the hexagonal-ice-like H-bonding structures can occur on (OH)$_2$-SAM. As shown in Fig. 1(c), it is demonstrated that the contact angles on $(OH)_2$-SAM are totally larger than 65° within the moderate range, which has been considered as a boundary of surface hydrophobicity and hydrophilicity [1,22,23].

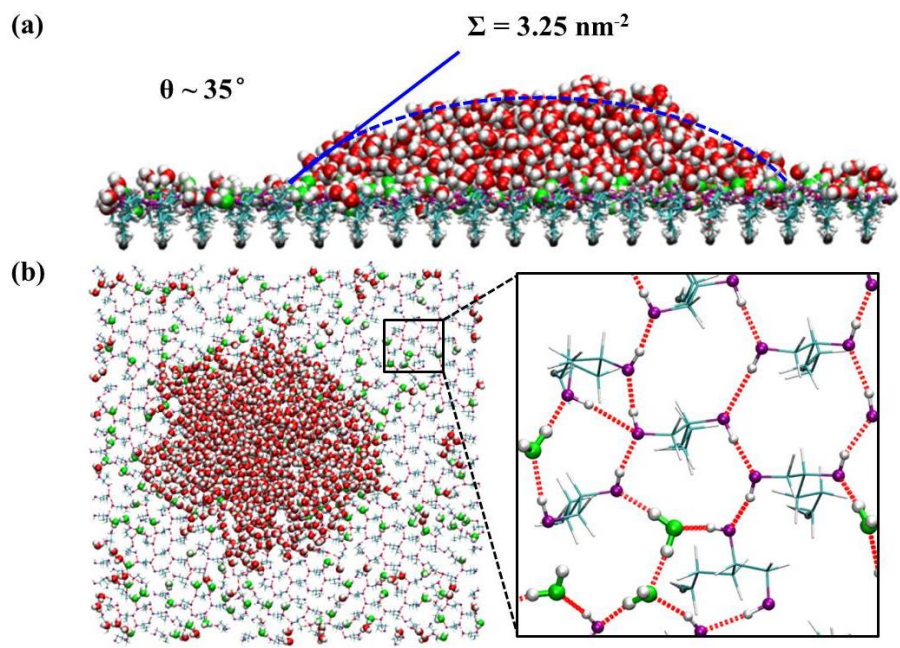

**FIG. 2.** (a) Side view snapshot of water droplet on $(OH)_2$-SAM in the loose range at $\Sigma = 3.25$ nm$^{-2}$ with water embedded. (b) Top view (left) of subfigure (right) with its close-up for water embedded structures. Atom representations and color settings are as in Fig. 1, except for embedded water in green and white balls.

With loosely packing densities ($\Sigma$) from ~2.5 nm$^{-2}$ to ~3.9 nm$^{-2}$, the loose hexagonal-ice-like structure can be still maintained, and the consequent hydrophobicity weakened to some extent was observed on $(OH)_2$-SAM. In Figure 1(c), water droplets still occur on $(OH)_2$-SAM, while their contact angles decrease gradually from ~65° to ~27° as $\Sigma$ decreases. Figure 2(a) and 2(b) show typical views of water droplet on $(OH)_2$-SAM with the packing density $\Sigma = 3.25$ nm$^{-2}$. We detected not only water droplets with the contact angle of 35°, but also water molecules outside of the droplet on $(OH)_2$-SAM surfaces (Fig. 2(b)). Extensive examinations showed that most of these outside water molecules simultaneously have two or three H-bonds with the OH groups in the OH matrix of $(OH)_2$-SAM and their z-coordinate not higher than 2 Å compared with the average oxygen position of the first hydroxy groups on $(OH)_2$-SAM. This type of water molecules was also found under the droplet, and their average energies with $(OH)_2$-SAM were -54.7 kJ/mol and their H-bond lifetimes were 141 ps (see PS5 in Supporting Information). We termed the water molecules having a water-SAM interaction energy $\leq$ -40 kJ/mol and an oxygen z-coordinate not higher than 2 Å compared with the average oxygen position of the first hydroxy groups on $(OH)_2$-SAM, as *embedded water*, and the composition that included both the embedded water and the OH matrix, was termed as *embedded-water-OH composite structure*. This is similar to our previous observation on the COOH-SAM where water is embedded into the COOH matrix to form embedded-water-COOH composite structure [16]. Herein, the embedded-water composite structure demonstrates again that "water is active" [25] to directly stabilize the whole H-bonding structure in the OH matrix and the embedded-water-OH composite structure takes shape.

Moreover, within the sparse range of $\Sigma$ (less than ~2.5 nm$^{-2}$), the contact angles of water droplets increase again on (OH)$_2$-SAM as $\Sigma$ decreases (see Fig. 1(c)). Careful examination showed the very sparse arrangements of OHs in the OH matrix cannot maintain the integrated H-bond network even under the help of water, and the OH groups prefer to form localized cyclic H-bonded structures to expose the hydrophobic tails (see Fig. S1 in Supporting Information). Meanwhile, within the dense range of $\Sigma$, the dense arrangement of OHs also breaks the hexagonal-ice-like H-bonding network in the OH matrix so that these OHs prefer to form H-bonds with water above. Thus, with high $\Sigma$, the (OH)$_2$-SAM becomes completely wet with zero contact angle (see Fig. S2 in Supporting Information).

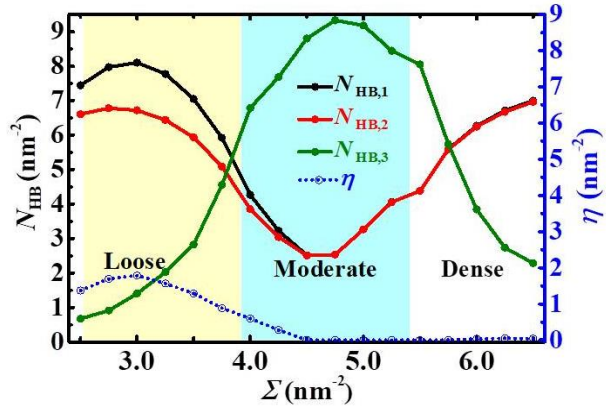

**FIG. 3.** Average numbers of H-bonds between the OH matrix of (OH)$_2$-SAM and water molecules within the region covered by water droplets ($N_{HB,1}$, black), between the embedded-water-OH composite structure and the water droplet above the composite structure ($N_{HB,2}$, red), and among the OH groups in the matrix under the water droplet ($N_{HB,3}$, green), together with the average numbers of embedded water molecules ($\eta$, blue, right blue axis) as a function of $\Sigma$.

Further, we show how the hydrophobicity occurs on this SAM surface terminated with originally hydrophilic OH groups. We have calculated the average numbers of H-bonds between the OH matrix and the water molecules within the region covered by water droplets ($N_{HB,1}$), between the embedded-water-OH composite structure and the water droplet above the composite structure ($N_{HB,2}$), and among the OH groups in the matrix under the water droplet ($N_{HB,3}$). Figure 3 illustrates $N_{HB,1}$, $N_{HB,2}$ and $N_{HB,3}$, as the function of $\Sigma$. In the moderate range of $\Sigma$ from 3.9 nm$^{-2}$ to 5.4 nm$^{-2}$, $N_{HB,3}$ has large values with its maximal value of 9.3 nm$^{-2}$, corresponding to the H-bond number in the intact hexagonal-ice-like structure where all OH groups are H-bonded with one another in the OH matrix. The H-bonds between the hexagonal-ice-like structure and the water droplet above, $N_{HB,1}$, has the much smaller values of ~2.5 – ~4.6 nm$^{-2}$, compared with the values of ~7.0 nm$^{-2}$ at $\Sigma$ = 6.5 nm$^{-2}$ for completely hydrophilic (OH)$_2$-SAM, indicating the very weak interaction between this SAM surface and water droplet above. This very weak interaction causes the hydrophobicity on (OH)$_2$-SAM. In the loose range of $\Sigma$, the numbers $\eta$ of embedded water molecules are ~0.7 – ~2.0 nm$^{-2}$; the H-bonds between the water-embedded-OH composite structure and the water droplet above, i.e., $N_{HB,2}$, has the values of ~4.1 – ~6.8 nm$^{-2}$, also less than $N_{HB,1}$ with the values of ~4.6 – ~8.1 nm$^{-2}$. The H-bond difference, ($N_{HB,2} - N_{HB,1}$), is up to ~1.4 nm$^{-2}$, indicating that the formation of water-embedded-OH composite structure really weakens the interaction between the composite structure and the water droplet above it, and consequently enhances the hydrophobicity of (OH)$_2$-SAM within this $\Sigma$ range compared with the super hydrophilicity in the dense range of $\Sigma$. The embedded-water-OH composite structure as well as the embedded-water-COOH composite structure in our previous work [16], both present the consistent physical mechanism for the enhancement effect of their hydrophobicity. Water droplets were stable on the ordered water monolayer on solid surfaces [26], but in our cases, the surface was not solid since the OH or COOH -terminated alkyl chains were flexible. Therefore, together with water-embedded-COOH composite, the water-embedded-OH composite demonstrate again that composite structures may exist on many other surfaces with hydrophilic groups, including previous observations on the coexistence of an ultrathin water layer and water droplets, e.g., on the surface of COOH-SAM [27] and the membrane formed with a bovine serum albumin -Na$_2$CO$_3$ mixture [28].

In summary, we have showed an unexpected hydrophobicity with a water contact angle of 82° on the flexible (OH)$_2$-SAM at room temperatures. This hydrophobicity attributes to the formation of a hexagonal-ice-like H-bonding structure in the OH matrix of (OH)$_2$-SAM, which sharply reduces the hydrogen bonds with water molecules above. Herein, this hexagonal-ice-like structure occurs on the (OH)$_2$-SAM within a wide range of the packing densities $\Sigma$, which presents not only the H-bonding oxygen-coordinate assignment consistent with the structure along the basal face of ice I$_h$ but also no dangling hydrogen atom out of the surface. To our best knowledge, this is for the first time to demonstrate the hydrophobicity on a flexible self-assembled surface terminated only with hydrophilic OH groups. Meanwhile, there remains the enhancement effect of hydrophobicity on the (OH)$_2$-SAM with the looser $\Sigma$, since water molecules can be embedded into the OH matrix to form embedded-water-OH composite structures as compensation for the loosened H-bonding structures and the integrated H-bonding network inside is maintained still.

We note that OH-SAMs continue to present puzzling observations after more than 30 years of studies, though generally terminated with single-OH groups. Experimental values of contact angles have ranged widely from 0° to 44°, which are collected from 39 research articles in literatures (listed in Table S1 of the Supplemental Material), while in previous theoretical calculations only the super-hydrophilicity on OH-SAM without any water droplet was indicated [3,4]. We think that these experimentally puzzling contact angles on OH-SAM can be understood in term of the water-embedded-OH composite structures with water droplets.

Meanwhile, we also note that interfacial wettability is of essential importance in extensive SAM-based applications [29,30], particularly on a variety of interacting processes ranging from biomolecules binding to cell adhesion [31-35]. Various complicated approaches have been applied to regulate or control the wettability, including the most popular way to mix terminal hydrophilic and hydrophobic groups onto SAMs [30,36-38], but may simultaneously cause intricate effects on other interfacial characteristics, such as adhesion, friction, conductivity, chemical activities and biocompatibility [3,34,37]. Here, the unique simple interface on the flexible SAM terminated only with two hydrophilic OH groups presents the unexpected hydrophobicity and provides a significant molecular-level platform for examining the bio-interfacial interactions ranging from biomolecules binding to cell adhesion. Potential applications might include various designs of anti-fouling, low-friction and thermal properties of solid/water interfaces.

The authors thank Jiajia Sun for preliminary testing calculations. This work was supported by the National Natural Science Foundation of China (Grants No. 12075201), Science and Technology Planning Project of Jiangsu Province (BK20201428), and the Special Program for Applied Research on Supercomputation of the NSFC-Guangdong Joint Fund (the second phase).

# Supporting Information for

# Unexpected Hydrophobicity on Self-Assembled Monolayers Terminated with Two Hydrophilic Hydroxyl Groups

**PS1. Literatures of experimental values of contact angles of water droplets on OH-SAMs**

**Table S1. Literatures of experimental values of contact angles of water droplets on OH-SAMs**

| Contact angle (°) | Contact angle (°) | Contact angle (°) | Contact angle (°) |
|---|---|---|---|
| ($\theta^a$)<10 [1] | ($\theta^a$)20±2, ($\theta^r$)11±3 [2] | ~20 [3] | 31.9±3.6 [4] |
| ($\theta^a$)<15 [5] | ($\theta^a$)25±2, ($\theta^r$)20±2 [6] | 22.9±3.0 [7] | 32 [8] |
| ($\theta^a$)10, ($\theta^r$)<10 [9] | ($\theta^a$)28±4, ($\theta^r$)19±4 [10] | 24.8±1.5 [11] | ($\theta^a$)~36, ($\theta^r$)~26 [12] |
| <10 [13] | 18±1.7 [14] | 25±3 [15] | 34±0.9 [16] |
| ($\theta^a$)<15 [17] | 18.1±1.6 [18] | 25 [19] | 38.2±0.32 [20] |
| <15 [21] | ($\theta^a$)~20 [22] | 25 [23] | 44±2 [24] |
| ~0 [25] | 16 [26] | 25.2±1.6 [27] | 4.6±1.6 [28] |
| 17±2.6 [29] | 29±3 [30] | 12±3 [31] | 17 [32] |
| 29.0±0.6 [33] | 13.5±0.5 [34] | 17.6±1.9 [35] | 29.4±1.6 [36] |
| ($\theta^a$) 16, ($\theta^r$)<5 [37] | 19 [38] | 31 [39] | |

**PS2. Snapshots of water covered on $(OH)_2$-SAM at $\Sigma = 2.0$ nm$^{-2}$ and 6.5 nm$^{-2}$**

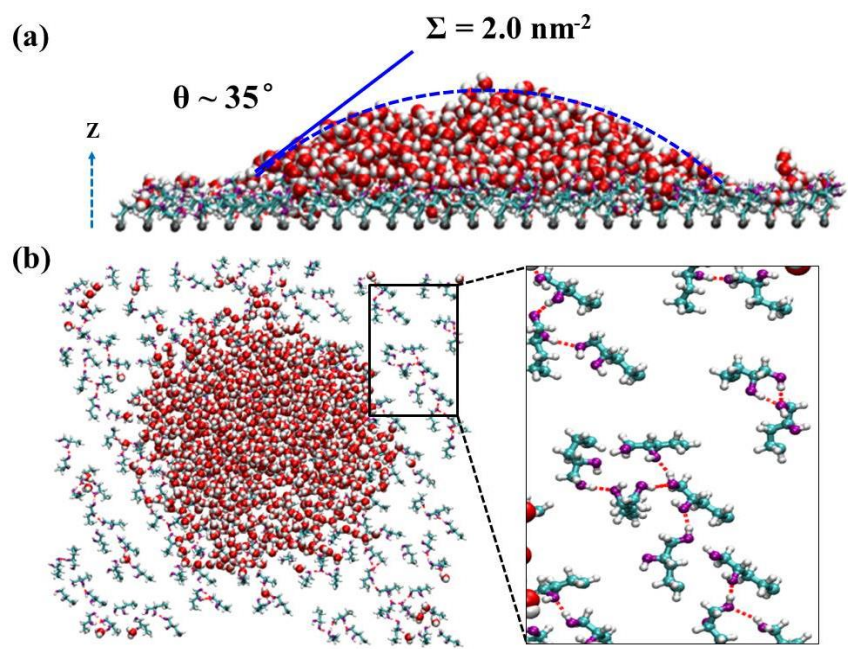

**Fig. S1.** (a) Side view snapshot of water droplet on $(OH)_2$-SAM at $\Sigma = 2.0$ nm$^{-2}$, (b) together with top view and its partial enlargement where OH groups prefer to form localized cyclic H-bonding structures with some exposure of the hydrophobic tails. Atom representations and color settings are as in Fig. 1.

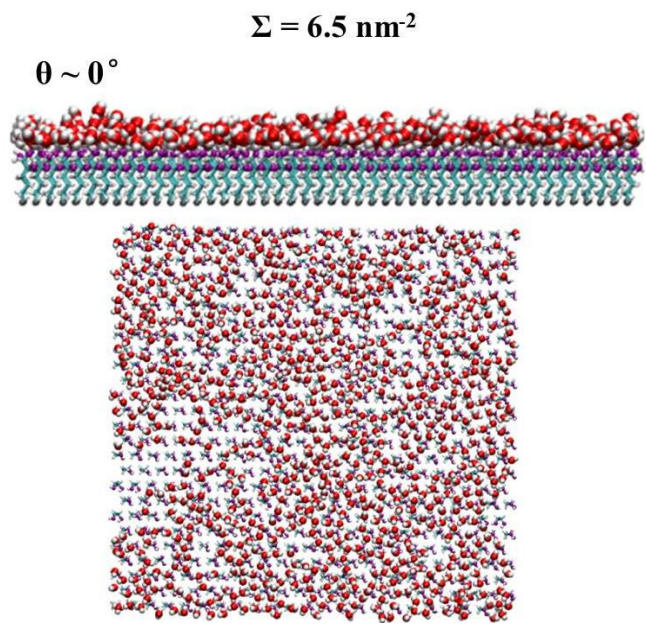

**Fig. S2.** Side view snapshot of water covered on $(OH)_2$-SAM at $\Sigma = 6.5$ nm$^{-2}$ together with top view on the bottom. Atom representations and color settings are as in Fig. 1.

## PS3. Method to compute the contact angles

The first water layer (0.35nm) above the $(OH)_2$-SAM is not considered. The number densities of water molecules in the droplets are estimated by cuboid lattices with dimensions of $0.05 \times 0.05 \times 0.6$ nm$^3$. Half of the number density of bulk water is used to estimate that of the gas-liquid boundary. The contact angles of water droplets are determined by fitting the number density distribution curves of the gas-liquid boundary to a circle.

## PS4. The interaction energies between water molecules outside of the droplets and $(OH)_2$-SAM

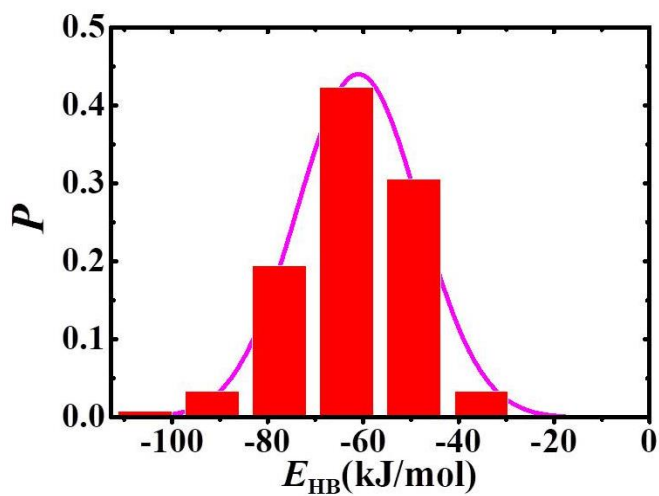

**Fig. S3.** The energies distributions between water molecules and $(OH)_2$-SAM outside of the droplets at $\Sigma = 3.25$ nm$^{-2}$.

## PS5. The calculation method of H-bond lifetimes

In this work, the H-bond lifetimes are characterized by the hydrogen bond autocorrelation function

$$C(t) = \frac{\langle h(0)h(t) \rangle}{\langle h(0)h(0) \rangle}$$

where $h(t) = 1$ if the tagged pair of atoms are continuously bonded from time 0 to time $t$, and $h(t) = 0$ otherwise. And $C(t)$ describes the probability a pair of atoms being bonded at time 0 and still bonded at time $t$.